%% file: apssamp.tex
\documentclass[%
 reprint,
 amsmath,amssymb,
 aps,
]{revtex4-2}

\usepackage{graphicx}% Include figure files
\usepackage{dcolumn}% Align table columns on decimal point
\usepackage{bm}% bold math
\usepackage[colorlinks,breaklinks]{hyperref}% add hypertext capabilities
\begin{document}

\preprint{APS/123-QED}

\title{Quantum Dynamical Microscopic Approach to Stellar Carbon Burning}% Force line breaks with \\
%%\thanks{A footnote to the article title}%

\author{Grant Close}
%% \altaffiliation[Also at ]{Physics Department, XYZ University.}%Lines break automatically or can be forced with \\
\author{Paul Stevenson}%
\author{Alexis Diaz-Torres}%
%% \email{Second.Author@institution.edu}
\affiliation{%
 Department of Physics, University of Surrey, Guildford, GU2 7XH, UK}%

\date{\today}% It is always \today, today,

\begin{abstract}
The process of carbon burning is vital to understanding late stage stellar evolution of massive stars and the conditions of certain supernovae. Carbon burning is a complex problem, involving quantum tunnelling and nuclear molecular states. Quantum dynamical calculations of carbon burning are presented, combining the time-dependent wave-packet method and the density-constrained time-dependent Hartree-Fock (DC-TDHF) approach. By limiting the contribution of triaxial molecular configurations to fusion, we demonstrate that the state-of-the-art DC-TDHF interaction potential successfully explains the appearance of some resonant structures in the sub-barrier fusion cross-section. With external perturbations to TDHF, we study the dynamic response of the compound nucleus to further explain resonant structure seen in the Gamow energy region. The results show the critical role of nucleon-nucleon interactions and compound nucleus excitations in the $^{12}$C + $^{12}$C fusion resonances observed at astrophysical energies.
\end{abstract}

\maketitle

%\tableofcontents

\input{Introduction.tex}

\input{Methodology.tex}

\input{Results.tex}

\input{Summary.tex}

\input{Appendix}

\bibliography{apssamp}% Produces the bibliography via BibTeX.

\end{document}

%% file: Introduction.tex
\textit{Introduction}-Understanding the dynamics of the $^{12}$C $+$ $^{12}$C reaction is an important factor in determining the path of stellar evolution, one that at astrophysical energies is not completely explained. Fusion cross-sections of the $^{12}$C $+$ $^{12}$C reaction determine the nucleosynthesis of heavier ions in carbon burning for stars with $M > 8 M_\odot$ and the conditions of type-1A supernovae \cite{Pignatari2013,Gasques2007,Wiescher2025}.

Direct measurements of the fusion cross-section have been made down to the center-of-mass energy $E_{c.m} = 2.1$ MeV, but struggle to probe further due to Coulomb effects \cite{Spillane2007,Fruet2020,Patterson1969,High1977,Aguilera2006,Mazarakis1973}. Measurements below the Coulomb barrier of the $^{12}$C $+$ $^{12}$C reaction, $E_{c.m} \approx 6$ MeV, have been achieved and presented in the form of the astrophysical S-factor, in which Coulomb effects are removed from the fusion excitation function. Resonant structures have been observed in the sub-barrier energy region which have been attributed to the formation of nuclear molecules in the compound nucleus $^{24}$Mg \cite{Bromley1960,Diaz-Torres2018,Taniguchi2021,Taniguchi2024}. Indirect experimental methods have been used to measure the S-factor down to the Gamow energy-peak for the $^{12}$C $+$ $^{12}$C reaction at $E_{c.m} = 1.5$ MeV. The indirect 'Trojan Horse Method' (THM) reveals many resonant structures in the S-factor in the Gamow energy region \cite{Tumino2018, Mukhamedzhanov2019}. Calculations based on antisymmetrised molecular dynamics predict some deep sub-barrier fusion resonances \cite{Taniguchi2024,Taniguchi2021}, which are consistent with the THM measurements \cite{Tumino2018}. Previous coupled-channels calculations had shown that the equatorial-equatorial orientation of the oblately (quadrupole) deformed $^{12}$C nuclei facilitates their capture \cite{Tanimura1978}, but did not show any resonant structures in the S-factor excitation function \cite{Assuncao2013,Jiang2013}. This has been demonstrated to be a result of not explicitly treating the role of specific alignments between the $^{12}$C nuclei in their fusion \cite{Diaz-Torres2018,Diaz-Torres2008}. The time-dependent wave-packet (TDWP) method has also been used to describe the $^{12}$C $+$ $^{12}$C fusion within a nuclear molecular picture. By implementing an alignment-dependent absorptive potential, the TDWP method shows that some resonant structures in the energy region around $3 \ \text{MeV} \leq E_{c.m} \leq 6 \ \text{MeV}$ can be explained \cite{Diaz-Torres2018}. Additional fusion resonances could be due to both cluster effects in the nuclear molecule \cite{Taniguchi2024, Diaz-Torres2024} and compound nucleus resonances \cite{Jiang2013}.

The current method builds upon the TDWP model, which solves the time-dependent Schrödinger equation with a collective Hamiltonian. Density-constrained time-dependent Hartree-Fock (DC-TDHF) theory is added to this model to produce ion-ion potentials to construct the potential operator. The extracted potential contains the dynamical aspects of the TDHF approach, including particle exchange, neck formation, nuclear deformations and excitations \cite{Umar2006,Simenel2018}. In contrast to dynamic DC-TDHF potentials, static density-constrained frozen Hartree-Fock (DC-FHF) potentials have been used in the literature \cite{Simenel2018}. The latter corresponds to a sudden (or diabatic) interaction that is physically realistic when the time scale of the internuclear radial motion is much shorter than the time scale of the single-particle motion. It is expected to be important at collision energies well above the nominal Coulomb barrier, but not at sub-barrier incident energies. The calculations of the present model demonstrate the crucial role of both triaxial nuclear molecular configurations and compound nucleus resonances in the formation of resonant structures in the astrophysical S-factor. We present the methodology of this approach followed by the results and summary.

%% file: Methodology.tex
\textit{TDWP method}-The dynamics of the collision are controlled by the TDWP method, which can be summarised into three main steps: (1) the definition of the initial wave-packet, i.e. $\Psi(0)$; (2) evolution of the wave-packet in time by solving the time-dependent Schr\"odinger equation; and (3) calculation of fusion probabilities and fusion cross-sections from the wave function, $\Psi(t)$, after the wave-packet has ceased interacting with the potential wells \cite{Vockerodt2019,Diaz-Torres2018}.

\begin{figure}[h]
    \centering
    \includegraphics[width=0.3\textwidth]{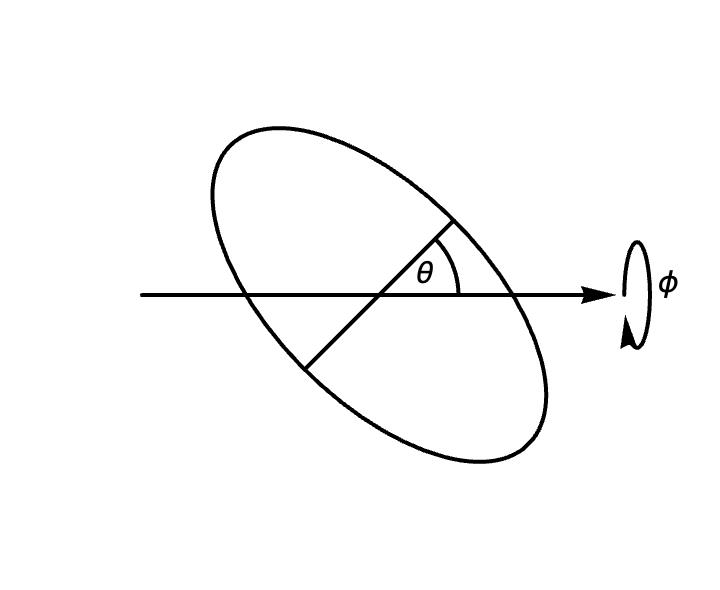}
    \caption{Diagram describing the definition of the angles $\theta$ and $\phi$ between the $^{12}$C symmetry (minor ellipsoid) axis with respect to the collision (horizontal) axis.}
    \label{fig:AngDiag}
\end{figure}

The dinuclear system is characterised by collective coordinates, which include the internuclear distance and orientation of the individual nuclei relative to the internuclear axis \cite{Diaz-Torres2018}. The initial wave function, describing both nuclei and their relative radial motion, is prepared with the nuclei sufficiently far apart, guaranteeing that there is no coupling between the $^{12}$C structure and the relative motion of the $^{12}$C nuclei. The $^{12}$C nuclei begin in their ground states, $j^{\pi} = 0^+$, and the initial wave function can be expressed as a product state:

\begin{equation}
\label{eq:Iniwave}
\Psi_0(R,\theta_1,k_1,\theta_2,k_2) = \chi_0(R) \, \psi_0(\theta_1,k_1,\theta_2,k_2),
\end{equation}
where $R$ is the internuclear distance, $\theta_i$ are the polar angles between the symmetry axis of the $i^{th}$ nuclei and the internuclear axis, and $k_i$ are the conjugate momenta of the azimuthal angles, $\phi_i$, as shown in Fig. \ref{fig:AngDiag}. 

The initial state $\chi_0$ is a boosted Gaussian wave-packet while the internal wave function $\psi_0$ is constructed from associate Legendre polynomials, the definition of which follows the study preceding this work \cite{Diaz-Torres2018}. For the explicit forms of the wave function, the corresponding kinetic energy operator, and the method of time propagation see Appendices \ref{WF}, \ref{Appen:Kinetic_Op}, and \ref{appen: ChebySehv}.

\textit{Fusion absorption-}To simulate fusion within the pole-pole potential well, an imaginary (absorptive) potential was implemented. This absorptive potential was described by a Woods-Saxon potential:

\begin{multline}
    \label{eq:ImagAb}
    W(R,\theta_1,\theta_2) = \left[ \frac{-V_{i}}{1 + \exp{(\frac{R - r_{i}}{a_{i}})}} \right] \cos{(\theta_1)} \, \cos{(\theta_2)},
\end{multline}
where the parameters $V_{i},r_{i},a_{i}$ are the strength, centre, and diffuseness of the Woods-Saxon potential respectively, and they are chosen to maximise the absorption when inside the potential well of the pole-pole orientation. This expression is physically motivated by Ref. \cite{Diaz-Torres2008}. The phenomenological term, $\cos{(\theta_1)}\cos{(\theta_2)}$, controls how strong the absorption is felt by the non-axially symmetric orientations.

\textit{Energy filtering}-Since the initial Gaussian wave-packet contains different translational energies, it remains to evaluate the wave function as a function of the resolved incident energies. This can be achieved via an energy projection method, and in this case the window operator is used \cite{Schafer1990}. The window operator is defined as 
\begin{equation}
    \label{eq:WinOP}
    \hat{\Delta}(E_k) \equiv \frac{\varepsilon^{2^n}}{(\hat{H}-E_k)^{2^n}-\varepsilon^{2^n}},
\end{equation}
with the energy resolution parameter, $\varepsilon$, and n is a positive integer. The parameter $n$ changes the shape of the energy bin. A value of $n =2$ gives the energy bin a Gaussian-like shape and was used in all simulations. This energy bin will be centred on the resolved energy, $E_k$. The expectation value of this operator, $\langle\Psi|\hat{\Delta}(E_k)|\Psi\rangle$, gives the probability that the system is in the evaluated state, $|\Psi\rangle$, which is found in the range $E_k \pm \varepsilon$ \cite{Schafer1991}. To include the radial and rotational effects of the kinetic energy operator the Hamiltonian was transformed into a matrix in the window operator \cite{Lee2022}. The transmission coefficients were calculated using the window operator by

\begin{equation}
    \label{TransCo}
    T(E_k) = 1 - \frac{\langle\Psi_f|\hat{\Delta}(E_k)|\Psi_f\rangle}{\langle\Psi_i|\hat{\Delta}(E_k)|\Psi_i\rangle},
\end{equation}
where $\Psi_f$ and $\Psi_i$ denote the final and initial wave-packets respectively.

\textit{DC-TDHF}-The potential energy operator is computed by extracting ion-ion potentials generated by the DC-TDHF method at specific orientations using the Sky3D code \cite{Umar2006,Maruhn2014,abhishek_tdhf_2024}. The main steps in the DC-TDHF method are: (1) Initiating a TDHF collision. (2) Pausing the reaction after a specified number of time steps to start a static HF calculation of the dinuclear system. The density operators of the nucleons are constrained to the values of the density operators when the TDHF collision is paused, 

\begin{equation}
    \label{eq:DC-TDHF}
    \delta\left\langle\hat{H} - \sum_{q=p,n} \int d\mathbf{r}\lambda_q(\mathbf{r})(\rho_q(\mathbf{r}) - \rho_q^{TDHF}(\mathbf{r}))\right\rangle = 0.
\end{equation}
The index $q$ is the nucleon species and $\lambda_q(\mathbf{r})$ are the Lagrange parameters constraining the proton and neutron densities at each point in space \cite{Umar2006,Simenel2018,Stevenson2019}. The static minimisation procedure ensures that the static energy computed is the value corresponding to the current stage of the TDHF dynamics. The minimised energy extracted is known as the density-constrained energy $E_{DC}(R)$. The interaction potential is then calculated via

\begin{equation}
    \label{eq:VDC}
    V_{DC}(R) = E_{DC}(R) - E_1 - E_2,
\end{equation}
where $E_1$ and $E_2$ are the static HF energies of the individual nuclei calculated separately from the DC-TDHF procedure. 
To compute the DC-TDHF $^{12}$C $+$ $^{12}$C potentials for different orientations, a means of rotating the nuclei is necessary. This was accomplished using active rotations on the single particle wave functions with the rotation operator parametrised by the Euler angles $\alpha,\beta,\gamma \in \mathbb{R}$ \cite{Pigg2014},
\begin{equation}
    \label{eq:Rot}
    R(\alpha,\beta,\gamma) = e^{-i\frac{\alpha\hat{J}_z}{\hbar}}e^{-i\frac{\beta\hat{J}_y}{\hbar}}e^{-i\frac{\gamma\hat{J}_z}{\hbar}}.
\end{equation}
See Appendix \ref{Appen:SPR} for a detailed description of the rotation on single-particle wave functions.

All $^{12}$C $+ ^{12}$C orientations considered were simulated with the parameters given in Table \ref{table:DC}. 

\begin{table}[h]
\centering
\caption{Parameter values for the DC-TDHF simulations.}
\label{table:DC}
\begin{tabular}{||c c c||} 
 \hline
 Variable & Value & Description \\ [0.5ex] 
 \hline\hline
 $x$ & $28$ fm &  Grid size in x-direction. \\ 
 $y$ & $28$ fm & Grid size in y-direction. \\
 $z$ & $56$ fm & Grid size in z-direction. \\
 $dx$ & 0.5 fm & Grid spacing in x-direction. \\
 $dy$ & 0.5 fm & Grid spacing in y-direction. \\ 
 $dz$ & 0.5 fm & Grid spacing in z-direction. \\
 $b$ & 0 fm & Impact parameter. \\
 $dt$ & 0.05 fm/c & Time step. \\
 $nt$ & 15000 & Number of time steps. \\
 Skyrme Force & SkI3 & Skyrme force parameter set. \\
 Pairing Force & None & Type of pairing force included. \\
 \hline
\end{tabular}
%\caption{Parameter values for the DC-TDHF simulations.}
%\label{table:DC}
\end{table}
The DC-TDHF simulations were performed using the SkI3 Skyrme parameter set \cite{Reinhard1995}. The choice of this set was due to the quadrupole deformation it produced. The static HF calculation for $^{12}$C gave $\beta_2 = 0.219$ and $\gamma = 60^\circ$, which is an oblate deformation. Experimental observations show that the $^{12}$C quadrupole deformation is approximately $\beta_2 = - \, 0.5$ \cite{Lebedev1999}. The static calculation was performed without applying a pairing force, as pairing effects are known to be negligible on fusion barriers and cross sections \cite{Ebata2014,Scamps2015}. In addition, Skyrme forces generally underpredict deformation for $^{12}$C, an effect which pairing worsens. Each simulation was initiated with the fragments separated  by $25 \ \text{fm}$, to ensure that the potential produced at this distance is solely described by the Coulomb potential. The initial centre of mass energy, $E_{c.m}$, was chosen to be close to the value of the peak of the Coulomb barrier. %The problem with choosing energy values too high is that the potential will start to incorporate dynamical effects that would not be present in the energies that are of interest in this study. This is an assumption that will need to be evaluated alongside experimental results, as we assume the dynamical effects seen at the Coulomb barrier or just above it are equivalent to the effects seen at sub-barrier energies \cite{Godbey2019}.
%The role of pairing in fusion proccesses in the TDHF formalism is currently unclear, with studies showing that the both the fusion barrier and cross-section are not effected by such considerations. As the inclusion of pairing would drive the $^{12}$C nucleus to become spherical it nas been neglected for the DC-TDHF calculations. 

\begin{figure}[h]
    \centering
    \includegraphics[width=0.48\textwidth]{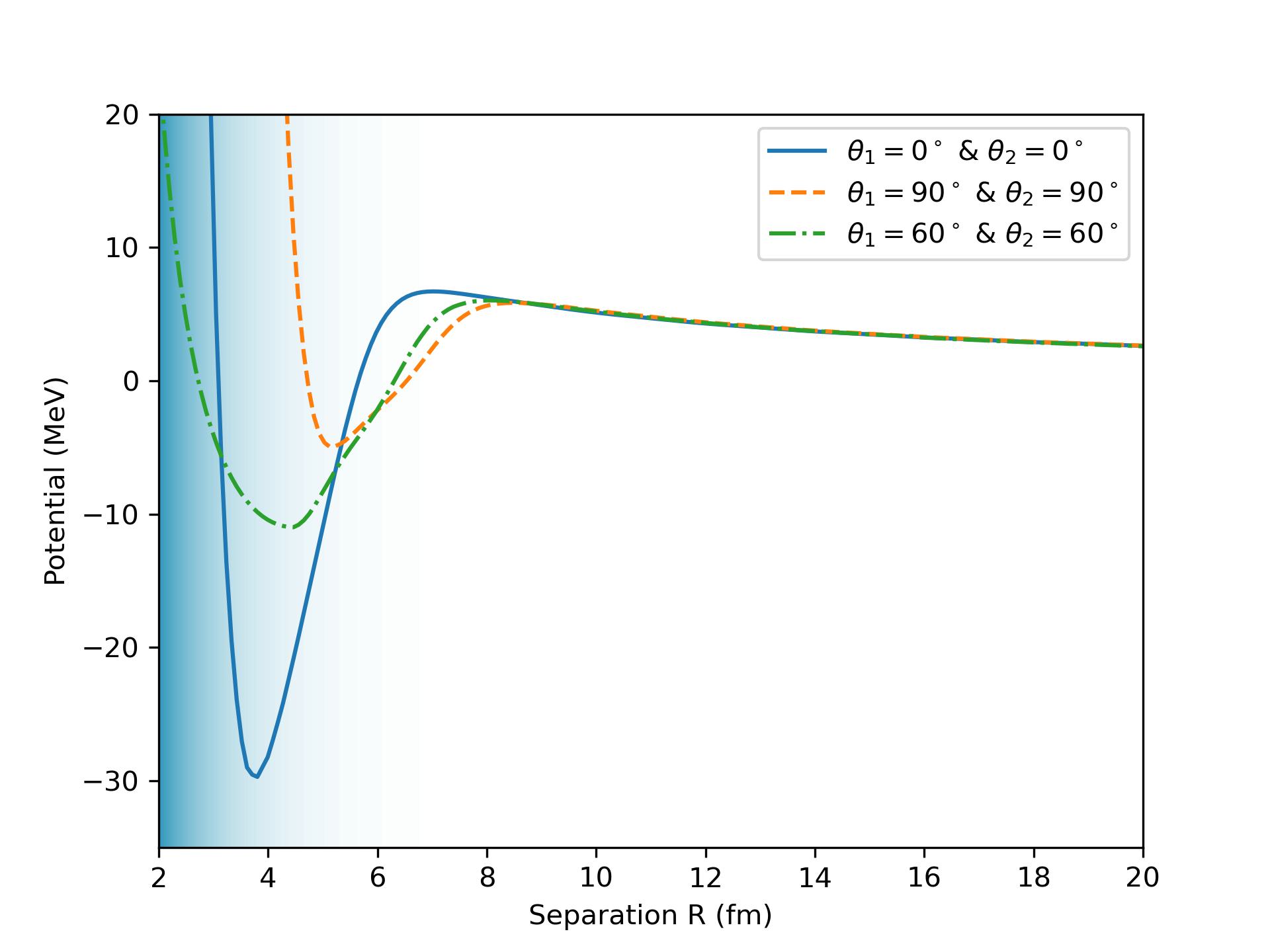}
    \caption{Fitted DC-TDHF potentials for the $^{12}$C $+ ^{12}$C reaction. With pole-pole, $(0^\circ,0^\circ)$, intermediate non-axially symmetric, $(60^\circ,60^\circ)$, and equatorial-equatorial, $(90^\circ,90^\circ)$ configurations. The active range of the pole-pole absorptive potential is represented by the shaded region, whose colour's intensity correlates with the absorption strength.}
    \label{fig:DC-TDHFPot}
\end{figure}

Using the single particle rotations in Eq. (\ref{eq:Rot}), all distinct ion-ion potentials were extracted from the $^{12}$C $+ ^{12}$C reaction, with potentials from equivalent orientations accounted for through the symmetry of the system. Asymptotically the DC-TDHF interactions are completely described by Coulomb forces and therefore can be extrapolated using the Coulomb potential. At short distances, the DC-TDHF potentials undergo a $\chi^2$ non-linear fitting routine to find the optimised parameters of a Woods-Saxon potential, which is only used to describe the repulsive hard core. Fig. \ref{fig:DC-TDHFPot} shows the interaction potential for 3 of the 55 calculated orientations. 

Fig. \ref{fig:DC-TDHFPot} shows the interaction potential for nuclei approaching in a non-axially symmetric orientation. In this configuration, the fraction of the wave-packet that tunnels through the barrier and enters the potential well will have 2 pathways. Either propagating back out of the well, due to the highly repulsive region ($R < 5 \ \text{fm}$), becoming a scattering state or contributing to fusion directly or/and via rearrangement of orientation. The location of the potential wells for the non-axially symmetric configurations are comparable to the pole-pole orientation. This could be a consequence of the static HF calculations producing a lower $\beta_2$ deformation than experimentally observed. Although, the height of the Coulomb barrier for such configurations indicates that a greater fraction of the wave-packet can tunnel through compared to the pole-pole orientation.

\textit{Compound Nucleus Resonances}- With the framework of TDHF it has been demonstrated that applying an external perturbation to the TDHF equations can reproduce giant resonances by performing a Fourier transform on the expectation value of the external perturbation function \cite{abhishek_tdhf_2024}. This can be used in this scenario by studying the dynamical response of the $^{24}$Mg compound nucleus to external perturbations using the SkI3 Skyrme parameter set. By investigating the excitation energy region in $^{24}$Mg in which the $^{12}$C + $^{12}$C reaction is energetically available (i.e., $13.9$ MeV above the $^{24}$Mg ground state), we can study excitations of the compound nucleus that can be occupied due to the reaction. Due to the identical nature of the system, monopole and quadrupole, $J = 0, 2$, calculations were carried out. Multipole calculations with $J=4$ were neglected as the TDWP fusion cross-section results for $J=4$ did not contribute enough in the region of interest, $E \le 3$ MeV. The available energy region lies in the beginning of the giant resonance and the resonant structures within can therefore be approximated with the Breit-Wigner formula,
\begin{equation}
    \label{eq:BW}
    \sigma_{BW} (E,J) = \frac{\pi\hbar^2}{\mu E} (2J+1)\frac{\Gamma_{in}\Gamma}{(E - E_r)^2+\frac{\Gamma^2}{4}}.
\end{equation}

 The entrance channel partial width, $\Gamma_{in}(J)$, is approximated to be $\Gamma_{in}(J) = T_J(E)\, \Gamma(J)$. The resonant structure are characterised by the FWHM and location of the peak, $\Gamma (J)$ and $E_r(J)$, respectively. The resonant structure extracted from the giant resonance calculations are parametrised in the form $(E_r,\Gamma)_{IS/IV}$, where $IS/IV$ denotes whether the excitation was isoscalar or isovector respectively. Isoscalar and isovector calculations describe if the protons and neutrons are moving in or out of phase relative to each other respectively. The monopole excitation results were $(2.24 \ \text{MeV}, 0.13\ \text{MeV})_{IS}$, $( 0.97\ \text{MeV}, 0.44\ \text{MeV})_{IV}$, $( 2.56\ \text{MeV}, 0.55\ \text{MeV})_{IV}$. The quadrupole results were $(2.23 \ \text{MeV}, 0.12\ \text{MeV})_{IS}$, $( 1.84\ \text{MeV}, 0.14\ \text{MeV})_{IV}$, $( 2.25\ \text{MeV}, 0.12\ \text{MeV})_{IV}$, $( 2.56\ \text{MeV}, 0.10\ \text{MeV})_{IV}$, $( 2.80\ \text{MeV}, 0.25\ \text{MeV})_{IV}$. In the quadrupole excitation, there seemed to be a resonance at the same location for the isoscalar and isovector calculations. This indicates that the resonant structure seen at this energy is not solely one or the other type. The contribution of the compound nucleus resonances to the total fusion cross section (indirect component) is given by $\sigma_{CN} = \sum_J \sigma_{BW}(E,J)$. 

%% file: Results.tex
\textit{Results and discussion}-All TDWP calculations used the parameters given in Table \ref{table:TDWP} to construct the collective five-dimensional space.

\begin{table}[h]
\centering
\caption{Parameter values for the TDWP simulations.}
\label{table:TDWP}
\begin{tabular}{||c c c||} 
 \hline
 Variable & Value & Description \\ [0.5ex] 
 \hline\hline
 $R_{min}$ & $0.5$ fm &  Minimum radial position. \\ 
 $R_{max}$ & $1000$ fm & Maximum radial position. \\
 $dR$ & $0.55$ fm& Radial grid spacing. \\
 $\sigma_0$ & 15 fm & Spatial width of Gaussian wave-packet. \\
 $R_0$ & $300$ fm & Centre of Gaussian wave-packet. \\ 
 $\Delta t$ & $10^{-22}$ s & Time step. \\
 $j_{max}$ & 4 & Max angular momentum of each nucleus. \\
 $k_{max}$ & 0 & Max projection of $j$. \\
 $V_i$ & 50 fm & Strength of absorptive potential. \\
 $a_i$ & 0.3 fm & Diffuseness of absorptive potential. \\[1ex]
 \hline
\end{tabular}
%\caption{Parameter values for the TDWP simulations.}
%\label{table:TDWP}
\end{table}

The angles $\theta_i$ and the conjugate momenta of the azimuthal angles, $k_i$, are determined by the KLEG-DVR method \cite{Sukiasyan2001,Otto2008,Otto2012}. The angles $\theta_i$ are discrete and have $j_i + 1$ evenly spaced values. $k_{max}$ are chosen for the orientation where the direction of spin is orthogonal to the internuclear axis, $k_i = 0$ (i.e., there is no reorientation of the $^{12}$C nuclei). The numerical error associated with both the norm of the wave packet and the conservation of energy is $\sim 10^{-12}$.

When generating the transmission probabilities for the $^{12}$C $+$ $^{12}$C reaction, one of the main features to consider is the position of the absorptive potential. Microscopic calculations in Ref.~\cite{Diaz-Torres2008} suggest that the orientation where fusion occurs is the pole-pole configuration, which is where the absorptive potential in Eq. (\ref{eq:ImagAb}) is centred. From Fig. \ref{fig:DC-TDHFPot}, it can be seen that the radial position of the potential pocket for non-axially symmetric configurations is similar to the pole-pole potential well. This feature is a consequence of the deformation value of $^{12}$C that was determined by the static HF calculation. Eq. (\ref{eq:ImagAb}) ensures that the non-axially symmetric orientations will undergo less fusion absorption than the pole-pole orientation.

\begin{figure}[h]
    \centering
    \includegraphics[width=0.48\textwidth]{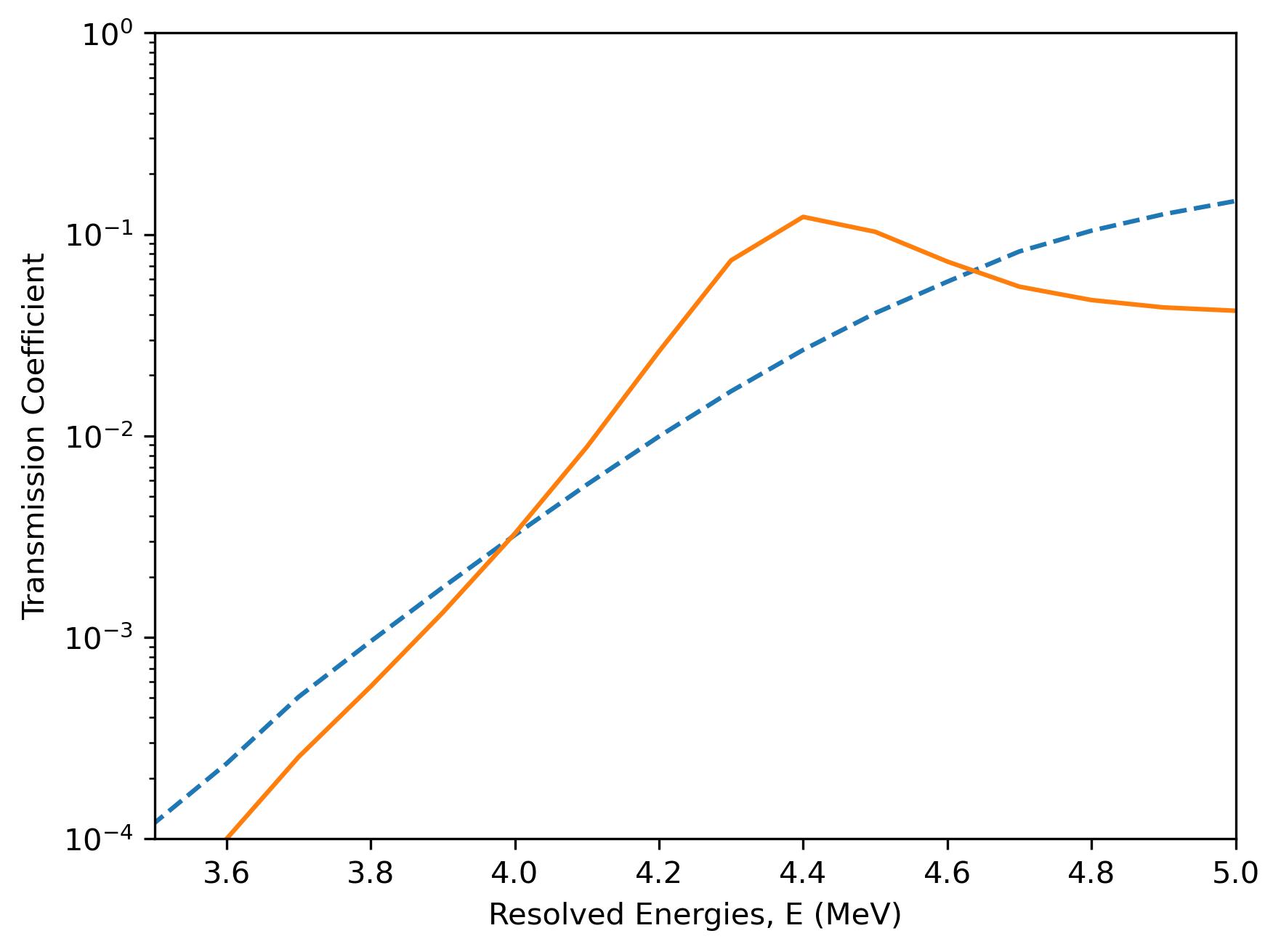}
    \caption{Direct transmission coefficients as a function of resolved incident energies for $^{12}$C $+ ^{12}$C head-on collisions. The curves differ by the shifting parameter  $r_{iw} = 0.0$ fm (blue dashed) and $r_{iw} = 0.70$ fm (orange).}
    \label{fig:ShiftAb}
\end{figure}

Fig. \ref{fig:ShiftAb} shows the transmission coefficients computed when the absorptive potential is centred at the minimum of the pole-pole potential (blue dashed line) for central collisions. There is no resonant structure because the non-axially symmetric orientations contribute strongly to the transmission coefficient. To further inhibit the absorption of the wave-packets in the non-axially symmetric configurations, the absorptive potential can be shifted towards the origin, by $r_i - r_{iw}$, so that the active region of $W(R,\theta_1,\theta_2)$, shaded area in Fig. \ref{fig:DC-TDHFPot}, is centred before the minimum of the potential pockets of the non-axially symmetric configurations. The wave-packet will then either propagate out of those potential pockets as a scattering state or contribute to fusion. Setting the parameter $r_{iw} = 0.70$ fm shifts the absorptive potential to the extent that the non-axially symmetric orientations largely do not directly contribute to fusion and a resonance peak is produced in the transmission coefficient, as shown in Fig. \ref{fig:ShiftAb} (orange solid line). 

\begin{figure}[h]
    \centering
    \includegraphics[width=0.48\textwidth]{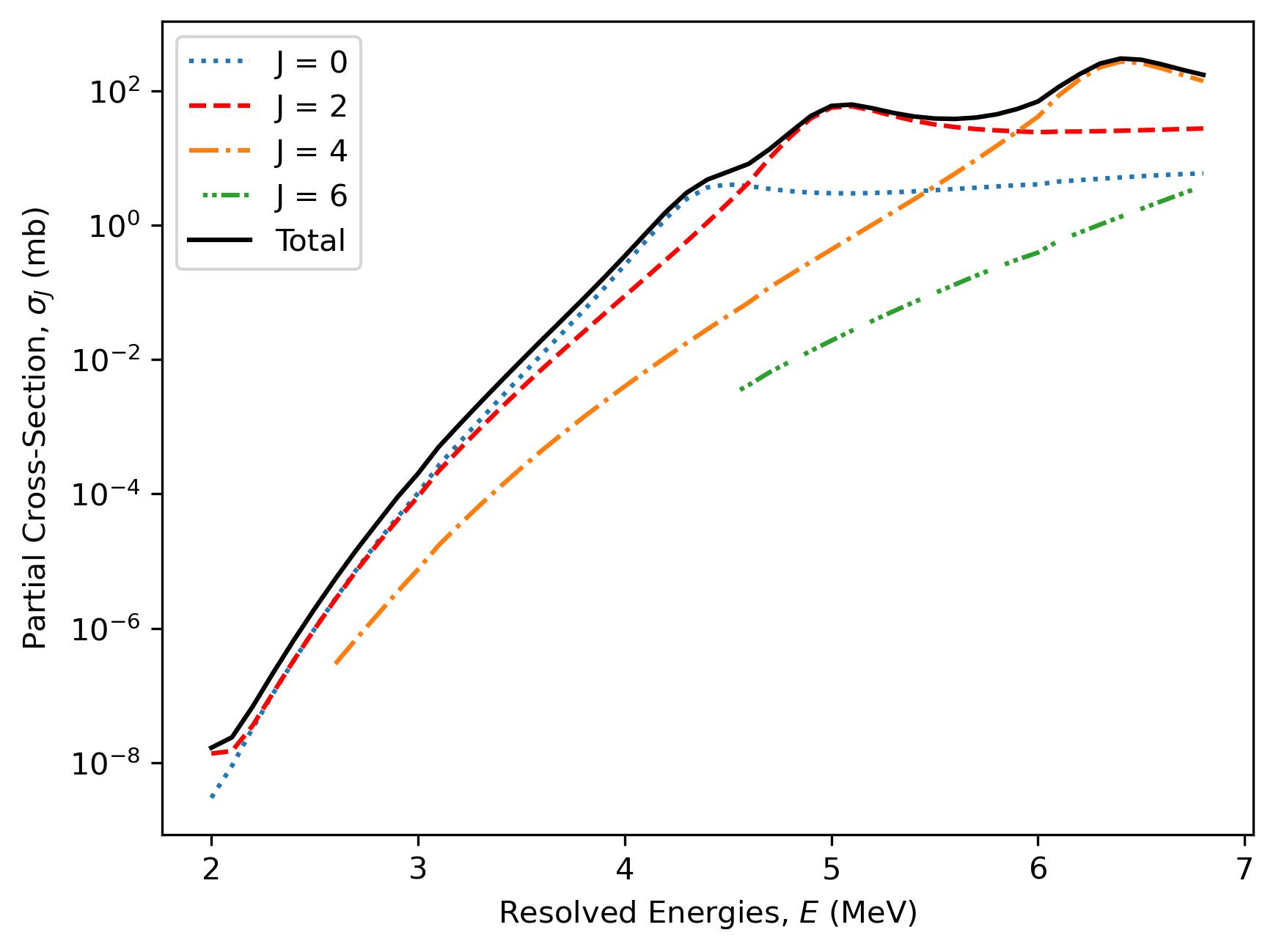}
    \caption{Direct component of the partial fusion excitation functions and their sum.}
    \label{fig:PartCross}
\end{figure}

Using the shifting parameter, $r_{iw} = 0.5$ fm, the TDWP calculations were performed for different values of the total angular momentum, $J$. For this reaction, the values of $J$ contributing to the low-energy region are $J = 0,2,4,6$. The direct component of the total fusion cross-section of the reaction was computed using $\sigma_{D} = \pi \hbar^2 (\mu E)^{-1} \sum_J (2J +1)T_J$, where $E$ is the incident centre-of-mass energy and $T_J$ is the partial transmission coefficient. The direct fusion cross-sections for the partial wave calculations are shown in Fig. \ref{fig:PartCross}. For the energy region that is below the Coulomb barrier of this reaction, the three resonant structures in the total fusion cross-section originate in the separate partial fusion excitation functions. At energies $E < 3.5$ MeV, the present model does not predict any additional resonant structure. We do not calculate energy values in the region $E_{c.m} < 2$ MeV, due to numerical instability arising from the method of calculating the transmission coefficient in Eq. (\ref{TransCo}). Higher values of $J$ will not contribute to the deep sub-barrier energy region of the total fusion cross-section.

\begin{figure}[h]
    \centering
\includegraphics[width=0.48\textwidth]{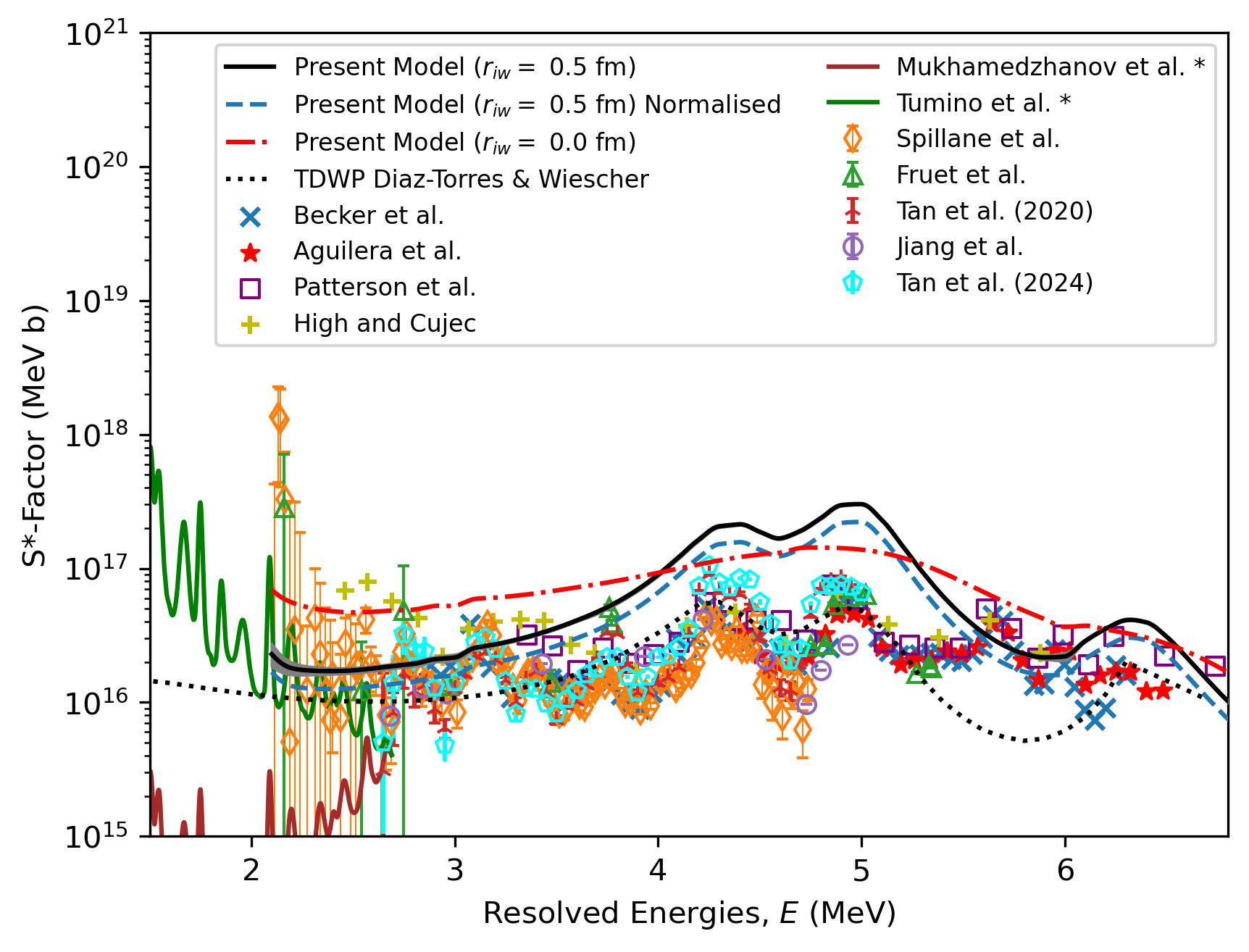}
        \caption{The modified astrophysical S-Factor for $^{12}$C $+ ^{12}$C as a function of the collision energy. The direct S*-factor function of the present model has a numerical error shown with grey shaded regions, which is due to the standard error in the mean values. The $*$ symbol in the legend denotes experimental data that has been renormalised to Tan et al. 2024.}
        \label{fig:Sfact}
 \hfill
\includegraphics[width=0.48\textwidth]{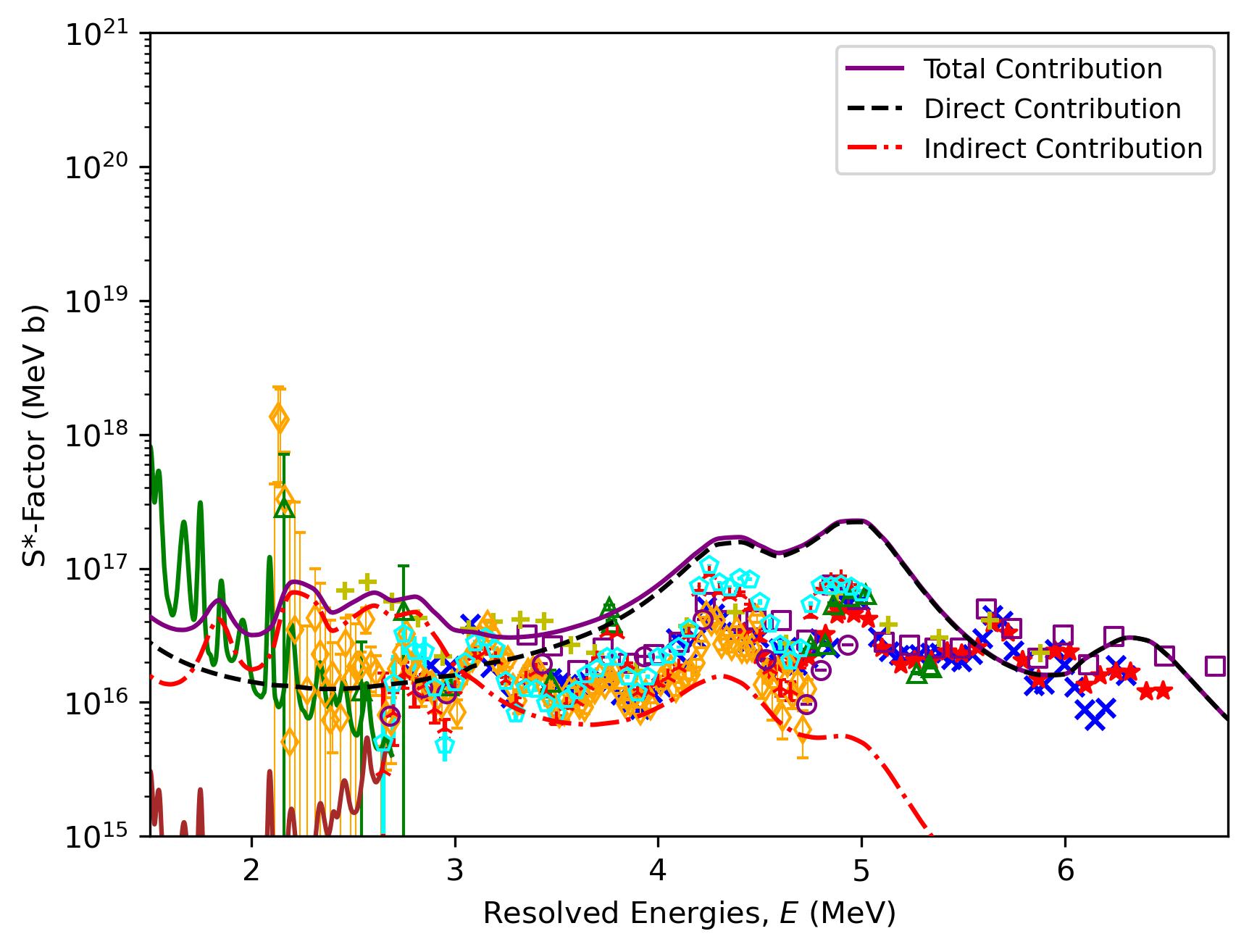}
        \caption{The same as in Fig. 5, but showing the decomposition of the total astrophysical S*-factor in its  direct and indirect components. Both components contribute to the formation of resonant structures in the total S*-factor.}
        \label{fig: TotalS}
\end{figure}

Fig. \ref{fig:Sfact} presents the modified astrophysical S-factor, $\textnormal{S*}(E)=\sigma_{fus}(E) \, E \, \exp(2\pi\eta+0.46 E)$, where $\eta$ denotes the Sommerfeld parameter, $\eta = (\mu/2)^{1/2}Z_1Z_2e^2/(\hbar^2E)^{1/2}$, and $Z_{1,2}=6$ for the charge number of $^{12}$C. In Fig. \ref{fig:Sfact}, the direct S*-factor of the present model is compared with previous TDWP results  and experimental data \cite{Spillane2007,Fruet2020,Patterson1969,High1977,Aguilera2006,Jiang2013,Tan2020,Tan2024,Becker1981,Diaz-Torres2018,Tumino2018,Mukhamedzhanov2019}. The $r_{iw} = 0.0$ fm curve over predicts the experimentally observed values. When compared to the curve for $r_{iw} = 0.5$ fm (black solid curve), the $r_{iw} = 0.5$ fm results seem to follow the same trend as the direct experimental results in the low energy region, $E < 3$ MeV but over predict the values for the resonant structure associated with $J = 0,2$. This is an indicator that the average height of the Coulomb barriers for all orientations is low, allowing too much of the wavepackets through. A possible cause of this discrepancy could be the relatively small oblate deformation value of the $^{12}$C nucleus ($\beta_2 = -0.219$) produced by the static HF calculation with the Skyrme force, SkI3. It causes the interaction potential for the non-axially symmetric orientations to be comparable to the potential for the pole-pole orientation, as shown in Fig. \ref{fig:DC-TDHFPot}. As the present approach uses a microscopic model to determine the interaction potentials, it does not include adjustable parameters to directly alter the curvature and depth of the potential pockets. This feature differs from the macroscopic, finite-range liquid-drop potential model employed in Ref. \cite{Diaz-Torres2018}, where the $^{12}$C quadrupole deformation considered is the experimental value ($\beta_2 = -0.5$). It shows the important role that nucleon-nucleon interactions have on the location of the resonant structure. As Ref. \cite{Diaz-Torres2018} used the experimental value for the deformation the heights of the Coulomb barriers are higher thus allowing less of the incident wavepacket to tunnel through the barrier. We normalise our results (blue dashed curve) using the transmission coefficient value of Ref. \cite{Diaz-Torres2018} at 2 MeV. At this energy, the transmission coefficient is almost entirely determined by the PP orientation, and the observed over prediction arises from the difference in Coulomb barrier heights. In Ref. \cite{Diaz-Torres2018}, the Hamiltonian operator used with the window operator in Eq. (\ref{eq:WinOP}) was an approximation of the full Hamiltonian (i.e., the asymptotic Hamiltonian that only accounts for the radial kinetic and potential energies). The present model includes all the terms of the kinetic energy operator, when the window operator method is used to calculate the transmission coefficients. Calculations showed that for $k_{max} = 0$ there was no difference in transmission coefficient between using the approximated and the full Hamiltonian in the window operator. The key differences between the present model calculations and those in Ref. \cite{Diaz-Torres2018} are (i) the use of microscopic interaction potentials, (ii) the use of the full Hamiltonian in the window operator method, and (iii) the inclusion of compound nucleus resonances. Using the results gathered by the giant resonance calculations we show the contributions to the fusion cross-section by simply adding the direct (TDWP) and indirect (GR) components, $\sigma_{fus} = \sigma_D + \sigma_{CN}$, shown in Fig. \ref{fig: TotalS}. The peak seen experimentally at $~2.2$ MeV can be explained by a culmination of monopole and quadrupole excitations that can be occupied at this energy.  Whereas, the monopole resonant peak seen at $2.5$ MeV could explain some resonant structure seen in experiments but due to the broadness of the curve it starts to add to the resonant peaks at lower energies. We show two resonant peaks below that seen by direct experimental results. Due to the method of generating our transmission coefficient we could calculate down to $2$ MeV before the results became unreliable, to probe down further we used a spline extrapolation method to $1.5$ MeV before this method broke down. We were therefore unable to show the S*-factor results of the monopole excitation resonant structure at $~1.0$ MeV, which is an important region due to the Gamow window.

%% file: Summary.tex
\textit{Conclusion}-In summary, the combination of DC-TDHF theory with the TDWP method has shown to be an effective means to describe the $^{12}$C + $^{12}$C fusion at stellar energies. It reproduces some fusion resonances at sub-barrier energies, and also produces the non-monotonic nature of the S*-factor in the astrophysically important energy region, $E_{c.m} < 3$ MeV. The appearance of some fusion resonances is due to the contribution made by triaxial nuclear molecular configurations, which resonate in their nuclear potential pockets, to the fusion cross section. Further improvements can be made to the present model calculations with the use of a different Skyrme force parameter set, which may predict better binding energies and quadrupole deformation for the $^{12}$C nucleus. Of all the parameter sets tested, SkI3 produced the quadrupole deformation that was most in agreement with experimental data, while some parameter sets resulted in a spherical $^{12}$C nucleus \cite{Desouza2024}. This could improve the agreement between the present model calculations and the experimental data. The framework provided by the TDHF method extended to study the compound nucleus resonances through giant resonance analysis has led to the identification of resonant structure within direct experimental measurements and predicted the existence of resonant structure beyond the current limit of direct experimental data. Progressing forward, the effects of clusters in the nuclear molecule and achieving stable low-energy direct results with the TDWP method will be investigated. %This could explore the origin of the fusion resonances revealed by the THM experiment for collision energies around the Gamow energy peak.

\textit{Acknowledgements-}This work was supported by the United Kingdom Science and Technology Facilities Council (STFC) under grants ST/X508810/1, ST/P005314/1, ST/V001108/1, and ST/Y000358/1. This work used the DiRAC Data Intensive service (DIaL2 / DIaL3) at the University of Leicester, managed by the University of Leicester Research Computing Service on behalf of the STFC DiRAC HPC Facility (www.dirac.ac.uk). The DiRAC service at Leicester was funded by BEIS, UKRI and STFC capital funding and STFC operations grants. DiRAC is part of the UKRI Digital Research Infrastructure. We thank Kyle Godbey for the addition of the DC-TDHF to the Sky3d code. For the purpose of Open Access, the author has applied a Creative Commons Attribution (CC BY) public copyright licence to any Author Accepted Manuscript version arising from this submission.

%% file: Appendix.tex
\appendix
\label{sec:appen}
\section{Wave Function}
\label{WF}
The $^{12}$C nuclei are prepared in their ground states, $j^{\pi} = 0^+$, and the initial total wave function can be expressed as a product state:

\begin{equation}
\label{eq:Iniwaveap}
\Psi_0(R,\theta_1,k_1,\theta_2,k_2) = \chi_0(R) \, \psi_0(\theta_1,k_1,\theta_2,k_2),
\end{equation}
where $R$ is the internuclear distance, $\theta_i$ are the polar angles between the symmetry axis of the $i^{th}$ nuclei and the internuclear axis, and $k_i$ are the conjugate momenta of the azimuthal angles, $\phi_i$.

The initial state $\chi_0(R)$ is a boosted Gaussian wave-packet 

\begin{equation}
\label{IniGauss}
\chi_0(R) = \mathcal{N}^{-1} \, \exp\left[\frac{-(R-R_0)^2}{2\sigma_0^2}\right]\, e^{-iK_0R},
\end{equation}
where $\mathcal{N}$ is a normalisation constant, and the parameters $R_0,\sigma_0, K_0$ are the initial centroid, spatial dispersion and the average wave number of the Gaussian wave-packet respectively. $K_0$ depends on the average incident energy $E_0$, $R_0$ and $\sigma_0$ and is found by solving $E_0 = \langle \Psi | \hat{H} | \Psi \rangle$ using the total Hamiltonian of the system, $\hat{H}$. The initial state dependent on the internal coordinates reads as

\begin{equation}
\begin{split}
\label{eq:Internalwave}
\psi_0(\theta_1,k_1,\theta_2,k_2) = [\eta_{j_1,m_1}(\theta_1,k_1) \, \eta_{j_2,m_2}(\theta_2,k_2)  
\\[1ex]
  + \, (-1)^J \, \eta_{j_2,-m_2}(\theta_1,k_1) \, \eta_{j_1,-m_1}(\theta_2,k_2) \, ] 
\\[1ex]
/\sqrt{2+2\delta_{j_1,j_2}\delta_{m1,-m2}},
\end{split}
\end{equation}

with

\begin{equation}
    \label{eq:LegPol}
    \eta_{j,m}(\theta,k) = \sqrt{\frac{(2j+1)(j-m)!}{2(j+m)!}}P^m_j(\cos\theta)\delta_{k,m}
\end{equation}
where $P^m_j(\cos\theta)\delta_{k,m}$ are associated Legendre functions. The initial angular momentum and projection quantum numbers for the individual nuclei are denoted by $j_i$ and $m_i$ respectively \cite{Sukiasyan2001,Otto2008,Otto2012}. In this case, $j_i=0$ and $m_i=0$ as the nuclei are prepared in their ground states. Physically, this sets the initial conditions to have an isotropic distribution of orientations. In Eq. (\ref{eq:Internalwave}), $J$ is the total angular momentum of the dinuclear system, which only takes even values due to the exchange symmetry of the system.
\section{Collective Kinetic Energy Operator}
\label{Appen:Kinetic_Op}
The reference frame in which the dinuclear system is described is the rotating centre-of-mass frame within a nuclear-molecule picture. The KLEG-DVR kinetic energy operator is expressed as
\begin{multline}
\label{eq:InternKO}
    \frac{2\hat{T}}{\hbar^2} = - \frac{1}{\mu}\frac{\partial^2}{\partial R^2} + \left(\frac{1}{I_1} +\frac{1}{\mu R^2}\right)\hat{j}^2_1 + \left(\frac{1}{I_2} +\frac{1}{\mu R^2}\right)\hat{j}^2_2 \\
    + \frac{1}{\mu R^2}[\hat{j}_{1+}\hat{j}_{2-} + \hat{j}_{1-}\hat{j}_{2+} + J(J+1) \\ 
    - 2k_1^2 -2k_1k_2 - 2k^2_2] - \frac{C_+(J,K)}{\mu R^2}\left(\hat{j}_{1+} + \hat{j}_{2+}\right) \\
    - \frac{C_-(J,K)}{\mu R^2}\left(\hat{j}_{1-} + \hat{j}_{2-}\right),
\end{multline}
with 

\begin{align}
\label{eq:CoriolisTerm}
    C_{\pm}(J,K) = \sqrt{J(J+1) - K(K \pm 1)}, \\ 
\label{eq:jsquared}
    \hat{j}_i^2 = -\frac{1}{\sin{\theta_i}}\frac{\partial}{\partial\theta_i}\sin{\theta_i} \frac{\partial}{\partial\theta_i} + \frac{k_i^2}{\sin^2{\theta_i}}, \\ 
\label{eq:jpm}
    \hat{j}_{i\pm} =\pm\frac{\partial}{\partial\theta_i} - k_i\cot{\theta_i},
\end{align}
where $I_i = 3B_0\beta_i^2$ is the rotational inertia of the nuclei, calculated with the deformation parameter $\beta_ = -0.219$ and $B_0 = 0.626 \, \hbar^2$ MeV$^{-1}$ \cite{Sukiasyan2001,Otto2008,Otto2012}. This value for the deformation parameter is approximately half experimentally observed values, but closest value achieved using a range of different Skyrme parameter sets in the HF calculations of the carbon nucleus. $J$ and $K$ are the total angular momentum of the system and its projection onto the internuclear axis, and the projection has discrete values ranging $[J,-J]$, defined by $K = k_1 + k_2$. The $\hat{j}_{i\pm}$ behave as ladder operators acting on the wavefunction with the value given in the $k\pm1$ component. The $C_{\pm}$ terms represent the Coriolis interaction throughout the dynamics of the reaction. This will change the $k_i$ conjugate momenta, altering the value of $K$, and will be neglected.
\section{Modified Chebyshev Propagator}
\label{appen: ChebySehv}
The Chebyshev propagator is a polynomial method aimed at expressing the time evolution operator as a sum of polynomial terms:
\begin{equation}
    \label{eq:cheb}
    e^{-i\frac{\hat{H}\Delta t}{\hbar}} \approx \sum_{n=0}a_nQ_n(\hat{H}_{norm}),
\end{equation}
where the argument, $\hat{H}_{norm}$, of the polynomial, $Q_n$, lies within the interval $[-1,1]$ \cite{Tannor}. To ensure that the normalised Hamiltonian values are within this interval it is defined by,

\begin{equation}
    \label{eq:Hnorm}
    \hat{H}_{norm} = \frac{\bar{H}\hat{1} - \hat{H}}{\Delta H}.
\end{equation}
The new terms are $\bar{H} = (\lambda_{max}+\lambda_{min})/2$ and $\Delta H = (\lambda_{max}-\lambda_{min})/2$, which incorporate the maximum and minimum eigenvalues of the Hamiltonian, $\lambda_{max}$ and $\lambda_{min}$ respectively, and the identity operator, $\hat{1}$. The Chebyshev coefficients, $a_n$, are computed to be 

\begin{equation}
    \label{eq:chebycoe}
    a_n = i^n(2-\delta_{n0})\exp\left(-i\frac{\bar{H}\Delta t}{\hbar}\right)J_n\left(\frac{\Delta H\Delta t}{\hbar}\right),
\end{equation}
the $J_n$ terms are the Bessel functions of the first kind. The polynomials $Q_n$, in Eq.(\ref{eq:cheb}), are recovered via a recursion relation with the initial conditions $Q_0(\hat{H}_{norm}) = \hat{1}$ and $Q_1(\hat{H}_{norm}) = \hat{H}_{norm}$, 

\begin{multline}
    \label{eq:recur}
    Q_{n-1}(\hat{H}_{norm}) + Q_{n+1}(\hat{H}_{norm}) \\
    - 2\hat{H}_{norm}Q_n(\hat{H}_{norm}) = 0.
\end{multline}
The inclusion of the imaginary absorption potential $\hat{W}$, to simulate fusion in the potential well, modifies the recursion relation in Eq.(\ref{eq:recur}). Leading to the modified Chebyshev propagator

\begin{multline}
\label{eq:modirecur}
 e^{-\hat{\gamma}} \, Q_{n-1}(\hat{H}_{norm}) + e^{\hat{\gamma}} \, Q_{n+1}(\hat{H}_{norm}) \\
    - 2 \, \hat{H}_{norm} \, Q_n(\hat{H}_{norm}) = 0.
\end{multline}
The initial conditions for the modified version become $Q_0(\hat{H}_{norm}) = \hat{1}$ and $Q_1(\hat{H}_{norm}) = e^{-\hat{\gamma}}\hat{H}_{norm}$. The operator $\hat{\gamma}$ links the Hamiltonian to the absorption potential via

\begin{equation}
\label{eq:abpot}
\hat{W} = \Delta H \, [\cos\xi \, (1 - \cosh \hat{\gamma}) - i\sin\xi \, \sinh\hat{\gamma}],
\end{equation}
where $\xi = \arccos(\frac{E-\bar{H}}{\Delta H})$ and $E$ is the mean collision energy \cite{Mandelshtam1995,2Mandelshtam1995}. For $\hat{W} = 0$, there is no absorption with $\hat{\gamma} = 0$ leading to the original recursion relation in Eq.(\ref{eq:recur}).
\section{Single Particle Rotations}
\label{Appen:SPR}
To compute the DC-TDHF $^{12}$C $+$ $^{12}$C potentials for different orientations, a means of rotating the nuclei is necessary. This was accomplished using active rotations on the single particle wavefunctions with the rotation operator parametrised by the Euler angles $\alpha,\beta,\gamma \in \mathbb{R}$ \cite{Pigg2014},
\begin{equation}
    \label{eq:Rotap}
    R(\alpha,\beta,\gamma) = e^{-i\frac{\alpha\hat{J}_z}{\hbar}}e^{-i\frac{\beta\hat{J}_y}{\hbar}}e^{-i\frac{\gamma\hat{J}_z}{\hbar}}.
\end{equation}
The total angular momentum operators, $\hat{J}_i$, operate around the $ith$ axis and when operating on single particle wavefunctions they have the form
\begin{equation}
    \label{eg:totang}
    \hat{J}_i = -i\hbar\hat{L}_i + \frac{\hbar}{2}\hat{\sigma}_i,
\end{equation}
with the Pauli matrices and the orbital angular momentum operator given by $\hat{\sigma}_i$ and $\hat{L}_i$ respectively. The exponential terms in Eq. (\ref{eq:Rotap}) can be approximated by a Taylor series expansion. In this model, the collision axis is set along the z-axis with the orthogonal axis being set to be the y-axis. For a rotation around the z-axis, Eq. (\ref{eq:Rotap}) takes the form,
\begin{equation}
    \label{eq:TayRot}
    |\psi^{\alpha + \Delta\alpha}\rangle = \sum^N_{n=0} \frac{(-i\Delta\alpha\hat{J}_z/\hbar)^n}{n!}|\psi^{\alpha}\rangle.
\end{equation}
Due to the inherent symmetry of the oblate $^{12}$C nucleus, the potentials produced by the $^{12}$C $+$ $^{12}$C are equivalent at certain orientations and therefore the amount of unique simulations needed are reduced.